# Data Augmentation Using Adversarial Training for Construction-Equipment Classification

Francis Baek, Somin Park, and Hyoungkwan Kim*

*Department of Civil and Environmental Engineering, Yonsei University, Seoul, Korea*

**Abstract:** *Deep learning-based construction-site image analysis has recently made great progress with regard to accuracy and speed, but it requires a large amount of data. Acquiring sufficient amount of labeled construction-image data is a prerequisite for deep learning-based construction-image recognition and requires considerable time and effort. In this paper, we propose a "data augmentation" scheme based on generative adversarial networks (GANs) for construction-equipment classification. The proposed method combines a GAN and additional "adversarial training" to stably perform "data augmentation" for construction equipment. The "data augmentation" was verified via binary classification experiments involving excavator images, and the average accuracy improvement was 4.094%. In the experiment, three image sizes (32-32-3, 64-64-3, and 128-128-3) and 120, 240, and 480 training samples were used to demonstrate the robustness of the proposed method. These results demonstrated that the proposed method can effectively and reliably generate construction-equipment images and train deep learning-based classifiers for construction equipment.*

## 1 INTRODUCTION

Advances in deep learning-based image analysis technologies make it possible to quickly and accurately obtain information from images, such as the location, status, and number of specific objects. In the construction industry, deep learning-based image analysis technologies such as object detection, tracking, and segmentation have been actively studied to automatically obtain information on construction resources, including workers, materials, and equipment (Luo et al., 2018), (Fang et al., 2018), (Son et al., 2019). The construction resource information can be applied to productivity analysis (Bügler et al., 2017), (Golparvar-Fard et al., 2012), (Kim et al., 2018); progress monitoring (Asadi et al., 2019), (Lei et al., 2019); and safety assessment (Fang et al., 2018), (Kolar et al., 2018), (Fang et al., 2018), (Fang et al., 2018), (Kim et al., 2019) on construction sites. These studies indicate the great potential of the deep learning-based approaches for significant improvements in construction management processes.

Deep learning-based image analysis for construction sites generally relies on supervised learning. Supervised learning, particularly for deep learning, requires a large amount of labeled images in which important elements of construction sites are marked. With the development of image devices, the task of collecting a sufficient amount of images has become easier. In particular, unmanned aerial vehicles improve the mobility of imaging devices, facilitating the acquisition of a sufficient amount of images both indoors and outdoors, as well as on large sites. However, despite the ease of acquiring a large amount of images, labeling the information in each image using bounding boxes or pixel-level segmentation requires considerable time and effort. In this context, preparing sufficient data for deep learning is still a challenging issue to be resolved.

To address the problem of data shortage, we propose the use of an adversarial training methodology in conjunction with generative adversarial networks (GANs) (Goodfellow et al., 2014) with the objective of data augmentation. Through this adversarial transformation process, the methodology deliberately generates and collects adversarial samples, which are used to improve the performance of a deep-learning model. The framework of the proposed method is presented in Figure 1. To the best of our knowledge, this is the first attempt to use a GAN coupled with adversarial training for a construction-resources classifier. The remainder of this paper is organized as follows. Comprehensive reviews on data augmentation are provided in Section 2. Section 3 explains the deep-learning structure and its components, which are described in Figure 1. In Section 4, descriptions of experiments, including nine cases, are presented, and the results are discussed. Finally, conclusions are drawn in Section 5.

*To whom correspondence should be addressed. E-mail: hyoungkwan@yonsei.ac.kr.

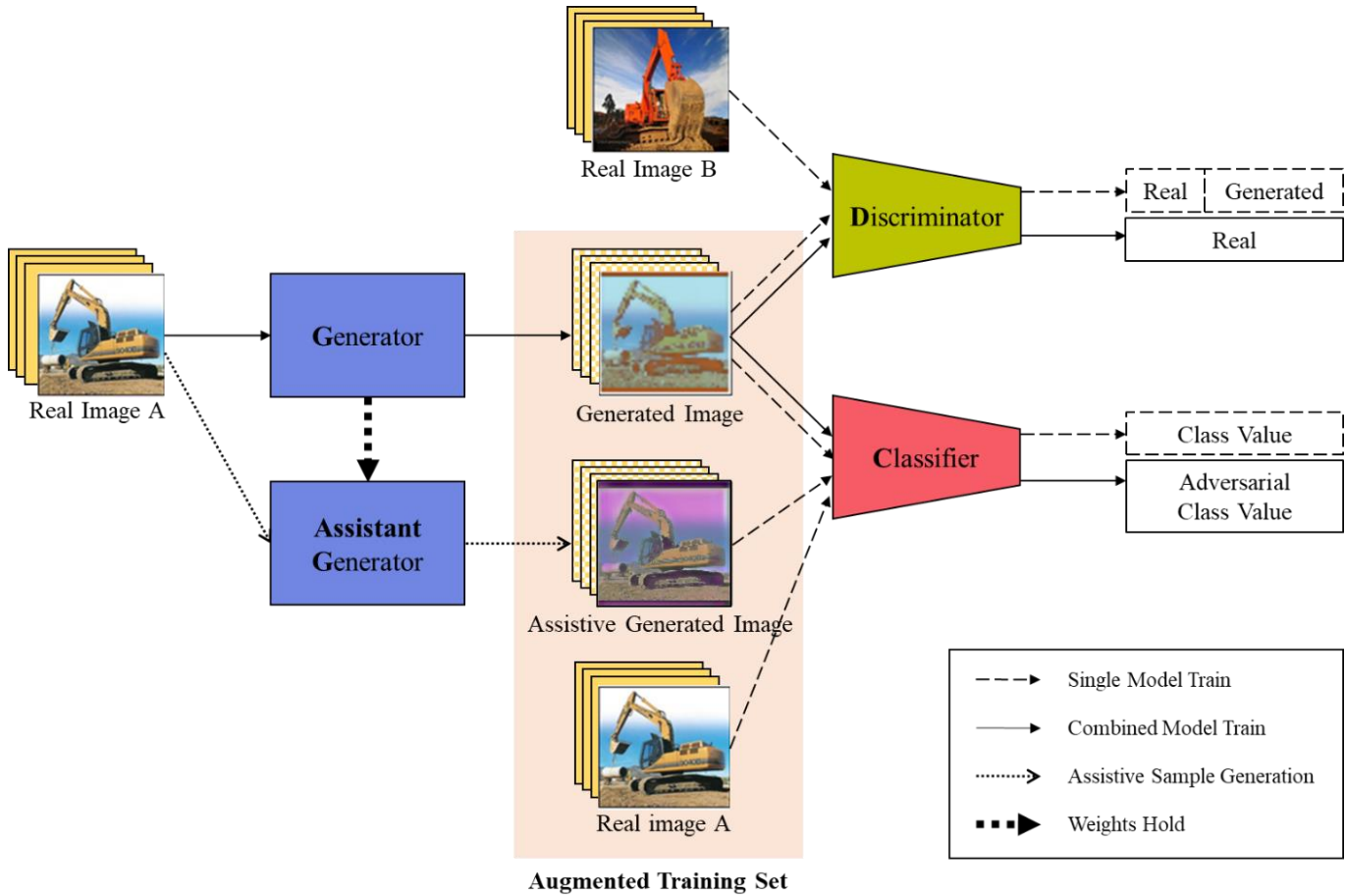

**Figure 1** Model structure.

## 2 RELATED WORKS

### 2.1 Deep learning-based image analysis in construction industry

Recently, studies have been actively conducted to manage infrastructure and construction sites through image analysis using deep learning. Field situations can be analyzed via the detection of workers (Fang et al., 2018), (Fang et al., 2018), (Fang et al., 2018), (Kim et al., 2019), (Fang et al., 2018), (Son et al., 2019), (Luo et al., 2019); construction materials (Fang et al., 2018), (Kolar et al., 2018), (Kim et al., 2019); and construction activities (Fang et al., 2018), (Luo et al., 2018). Studies have been conducted to identify various types of damage and defects such as cracks, pop-outs, and spalling (Zhang et al., 2019), (Bang et al., 2019), (Park et al., 2019), (Cha et al., 2017), (Wang and Cheng, 2019), (Wang et al., 2018), (Gao and Mosalam, 2018). Additionally, deep learning-based image analysis has been used for monitoring different types of infrastructure, such as roads (Bang et al., 2019), (Park et al., 2019); pipes (Wang and Cheng, 2019); and concrete bridges (Zhang et al., 2019), for timely and cost-effective maintenance of the infrastructure. These studies have revealed the potential of deep learning for the construction industry and confirmed the need for a large amount of construction-image data.

### 2.2 Simple image manipulation

Various data-augmentation techniques have been investigated for transforming images into other types of images and using them as if they were new data. Geometric transformation techniques such as flipping, cropping, rotation, and translation can artificially modify the geometry, position, and shape features of objects in an image with simple operations, allowing the generation of new features of objects. In addition to geometric transformation, changing the lighting conditions and sharpness of images can artificially produce new features of objects. These methods have the advantage that the data augmentation can be performed with simple image manipulations. However, the disadvantage is that human inspection is required, because changes in geometry or lighting can cause the objects in an image to lose their original features (Shorten and Khoshgoftaar, 2019).

### 2.3 Multiple image mixing

Attempts have been made to mix two or more images using the arithmetic mean value (Inoue, 2018) or a nonlinear method (Summers and Dinneen, 2019), for the purpose of data augmentation. While these methods had the effect of data augmentation, the mixing outputs were in some cases subject to human interpretation (Shorten and Khoshgoftaar, 2019). The cut-and-paste method, which involves cutting an object using a bounding box (Rao and Zhang, 2017) or segmentation (Dwibedi et al., 2017) label and pasting it into another image, has the advantage of creating a synthetic image with labels. However, the cut-and-paste method has a limitation: it cannot create new information that is not included in the given data, because it cuts and pastes the object within the given data.

Neural style transfer (Gatys et al., 2015) or CycleGAN (Zhu et al., 2017) is a technique that changes the image style while maintaining the contents of an image. Depending on the style of interest, an object can be expressed in new ways by expressing the object with various styles. However, this method has the disadvantage of having to collect images with the style of interest.

### 2.4 Image generation

Techniques for generating new data corresponding to objects rather than transforming data are also being studied. After a three-dimensional (3D) model based on an object is generated, the 3D model can be photographed from various angles to generate new features of the object (Kim and Kim, 2018). Such methods can automatically create new perspective or posture information for objects. However, in contrast to other image manipulation techniques, a 3D model construction process is necessary, which is an extra procedure. The disadvantage is that if a large number of objects are to be dealt with, as many 3D models as the number of objects are required. Since the advent of the GAN (Goodfellow et al., 2014), variations such as DCGAN (Radford et al., 2015), PGGAN (Karras et al., 2017), and BigGAN (Brock et al., 2018), have been proposed to create fake images (data augmentation). The use of GANs is increasing because of the ability to create a variety of new virtual images. However, training GANs requires a large amount of data, and the training process is unstable (Radford et al., 2015), (Shorten and Khoshgoftaar, 2019). In addition, if the fake image does not completely reproduce the object of interest, the original features can be lost; the classification accuracy may be lower with the use of augmented data than with the use of only real data (Ravuri and Vinyals, 2019).

### 2.5 Augmentation optimization

Even though there are various data-augmentation methods, such as image transformation and style transfer, applying all these methodologies does not always ensure the optimal performance and is not practical with regard to computation. In addition, even if one method is selected, it may be difficult to select an optimal value when the number of parameters to be selected is large, such as the angle of rotation or the degree of sharpness. Among the many methods, unnecessary augmentation schemes can be filtered out, and the results can be efficiently improved if one can find augmentation schemes and effective parameter values that are suitable for each problem.

Meta-learning uses other deep-learning or neural-network models to improve the performance of deep-learning models. The term meta-learning means "learn to learn." As the meaning of the term suggests, data-augmentation methods using meta-learning employ neural networks to train the model to find the best data-augmentation solution for the problem. Perez and Wang (2017) showed how to synthesize two images using a neural-network model for data augmentation, and Lemley et al. (2017) used more than one neural network to merge two or more images for augmented data. To find the optimal combination of schemes, such as sharpening and rotating, Cubuk et al. (2018) used the concept of reinforcement learning, and Mania et al. (2018) presented a model-free random search algorithm. These meta-learning methods are expected to achieve high performance because they attempt to find the optimal settings for augmentation of the scheme parameter values and neural-network weights.

Adversarial training is another way for data augmentation to generate more robust models (Goodfellow et al., 2015), (Li et al., 2018), (Samangouei et al., 2018), (Lee et al., 2017). Adversarial attacking approaches (Moosavi-Dezfooli et al., 2016), (Su et al., 2019), (Engstrom et al., 2017) show that neural networks can be fooled by simple methods such as noise injection, pixel-value changes, and simple transformations such as rotations and translations. Adversarial attacking aims at deceiving a deep-learning model and intentionally transforms the data so that the model cannot correctly analyze it. Such transformed data can be used as training data to compensate for the weaknesses of the deep-learning model, contributing to training more robust models; this concept is referred to as "adversarial training."

Herein, we propose adversarial training using a GAN as a data-augmentation method for construction equipment. The proposed methodology jointly trains a classifier with a generator. This approach has been used in other studies (Mounsaveng et al., 2019), (Lee and Seok, 2019), (Bousmalis et al., 2017). The novelty of the proposed methodology is that it is the first attempt to apply adversarial training to construction-equipment analysis. The study introduces a novel training procedure, which uses an assistant generator to support the classifier training. Additionally, the methodology was evaluated using various sizes of input images and different number of samples, validating the robustness of the data-augmentation capability of the proposed method.

## 3 METHODOLOGY

Adversarial training is one of the ways to identify and compensate for the weak points of a deep-learning model. As shown in Figure 1, the model for adversarial training used in this study comprised a generator, a discriminator, and a deep-learning model (classifier) to improve the performance. The objective of this study was to improve the performance of the binary classifier for excavator images. The generator, discriminator, and classifier augment the data by forming two independent combined models: a generator–discriminator combined model and a generator–classifier combined model. Both combined models intentionally use the original training sample to produce adversarial samples that can reduce the classification confidence of the classifier. The adversarial samples, along with the original training samples, are then used to train the classifier, allowing the classifier to improve the weak classification boundaries that are vulnerable to the adversarial samples. This methodology aims to generate a variety of adversarial samples through the interaction of combined models to train the classifier for more robust performance.

### 3.1 Generator–discriminator combined model

The combined model of the generator and discriminator in Figure 1 has the same objective as the general GAN structure of Goodfellow et al. (2014). The GAN consists of a generator and a discriminator and typically learns through alternation of the two processes. The discriminator learns to distinguish between real images and generated images. The ultimate purpose of the combined model is to train the generator. In the learning process of the generator, the weights of the discriminator of the combined model are fixed to be untrainable so that only the weights of the generator can be trained. The combined model is trained to produce output images that are to be classified as "real." Therefore, for the combined model to minimize the value of its loss function, the generator must be able to fool the discriminator into misclassifying the generated images as "real" images.

As shown in Figure 2, the generator creates a new image by performing pixelwise multiplication of the input image and the generated mask. The generator is designed based on U-Net (Ronneberger et al., 2015), which exhibits outstanding performance in medical-image segmentation. In contrast, the discriminator has a regularization purpose to prevent the generated images from being far outside of the domain of the real image. In this study, the real image domain refers to the image domain corresponding to the images of the validation set. As shown in Figure 3, the discriminator consists of general convolutional layers and has a patch output for evaluating the image in small regions. Because the image is analyzed by the patch units instead of as a whole, the generator must generate sophisticated images to fool such discriminators. Consequently, the generator produces sharper and higher-quality images than it does when evaluating the entire image at once (Isola et al., 2017). In this study, the objective function of the GAN used the least-squares loss, as follows:

$$\min_G \max_D L_{G-D}(G, D) = E_X[(1 - D(X))^2 + D(G(X))^2], \quad (1)$$

where *D* and *G* represent the discriminator and generator, respectively.

### 3.2 Generator–classifier combined model

A binary classifier was used to verify the data-augmentation performance of the adversarial training. In adversarial training, adversarial samples are needed to fool the classifier. To generate these adversarial samples, the proposed methodology uses a combined model of a generator and a classifier. Similar to the GAN training process described in Section 3.1, the generator and classifier are trained by alternating the two processes. First, the classifier is trained using the augmented training samples consisting of the original samples and the generated samples. In this study, the class of the excavator was represented by 1.0, and the other class was represented by 0.0. The classifier and the generator form the combined model. When the generator is trained, the weights of the classifier are fixed so that the combined model only trains the generator. As such, the combined model is trained to produce the output image belonging to the adversarial class, which differs from the original true class. To this end, the target value of the excavator class was adjusted to 0.8, whereas the other target value was adjusted to 0.2. For minimizing the loss function of the combined model, the output of the classifier analyzing the generated image should be close to the adversarial-class values. Thus, the generator learns to generate an image to intentionally reduce the classification confidence of the classifier from 1.0 to 0.8 (or from 0.0 to 0.2). The generated image is used in the classifier learning process along with the original training sample. Thus, the classifier becomes a more robust model by addressing the issue of weak classification boundaries based on the training using the adversarial samples. The loss function of the generator–classifier network is given as follows:

$$\min_{G,C} L_{G-C}(G, C) = E_{X,Y}[-Y\log C(X) - (1-Y)\log(1-C(X))] + E_{X,Z}\left[\left(Z - C(G(X))\right)^2\right], \quad (2)$$

where $C$ and $G$ represent the classifier and generator, respectively, and $Y$ and $Z$ correspond to the true class value and the adversarial-class value, respectively.

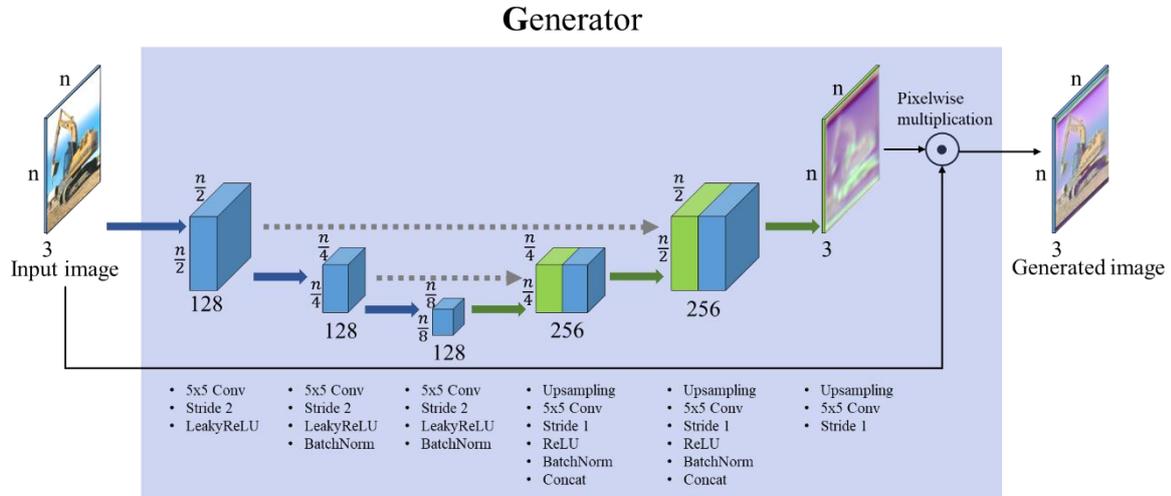

**Figure 2** Generator structure.

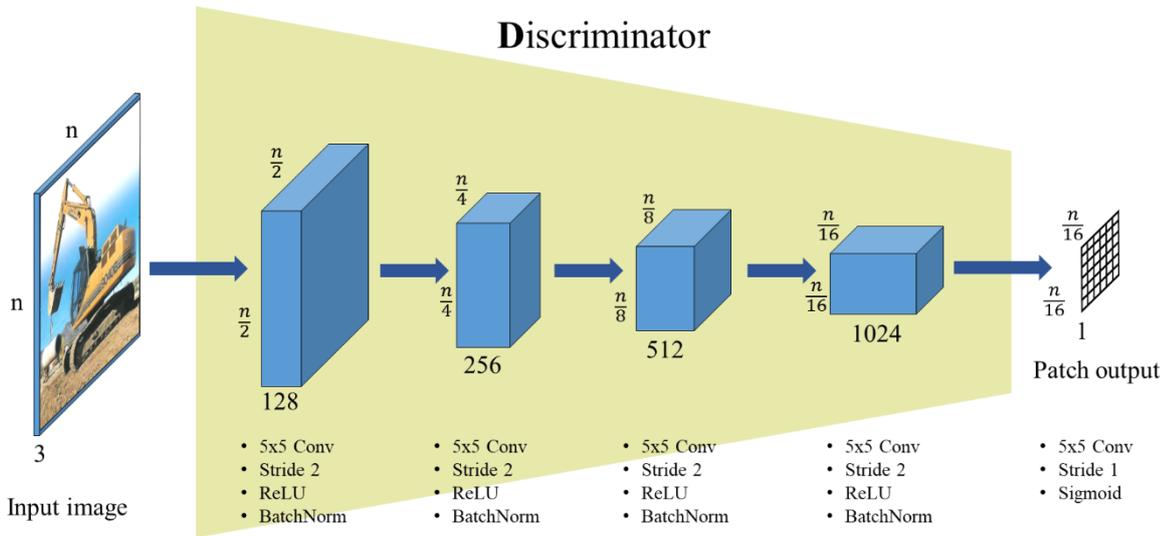

**Figure 3** Discriminator structure.

### 3.3 Entire model training using assistant generator

Sections 3.1 and 3.2 describe the training process of the two combined models in Figure 1. The GAN transforms the input image while satisfying the condition that the generated image does not deviate significantly from the real image domain. The combined model of the classifier and generator generates an image. This can present a challenging classification problem for the classifier. In this study, the two combined models independently update the weights of the generator. The discriminator is trained to distinguish between generated images and real images in the validation set. The discriminator competes against the generator to become intelligent enough to distinguish generated images from real images, while the generator has the purpose of creating fake images that appear authentic enough to fool the discriminator. The classifier is trained using both the original training samples and the generated training samples. The original training samples includes both positive (excavator) and negative (non-excavator) sets. As shown in Figure 1, the generated training samples consist of two sets: generated images and assistive generated images. The assistant generator has the same structure as the generator. The weights of the assistant generator are updated to the weights of the generator whenever the classifier achieves the best validation accuracy. The purpose of the assistant generator is to assist the training of the classifier by keeping the best weights that have been obtained until the time of iteration for the highest validation accuracy. It is possible that the quality of the generated images would rather decrease during training if the classifier or discriminator overwhelms the generator. In other words, if the classifier or discriminator becomes much more intelligent than the generator during training, the generator can temporarily lose the direction of training. In this case, the assistant generator continues to supply images generated by the best weights, preventing the generator from converging to the generation of low-quality images. Both the generator and assistant generator produce images corresponding to the adversarial-class value. Therefore, the methodology has the effect of tripling the number of training samples. Thus, the classifier is trained to be a part of the generator–classifier combined model. The two combined models are alternated, resulting in the proposed adversarial model, to generate new training samples in every iteration and train the classifier in an end-to-end manner.

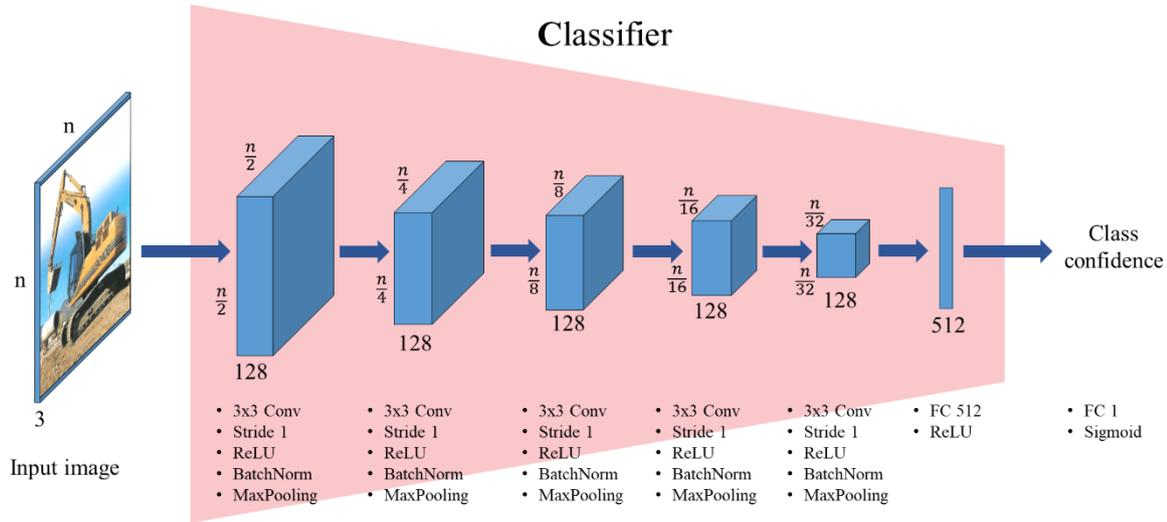

**Figure 4** Structure of classifier.

## 4 EXPERIMENTAL STUDY

### 4.1 Experimental setup

For the evaluation of the proposed methodology, training samples were collected from the ImageNet Large Scale Visual Detection Challenge dataset (Russakovsky et al., 2015). Excavator images were used as positive samples, and the negative samples comprised dump truck, mixer, and dozer images. The number of positive and negative samples were identical, and among the negative samples, the numbers of images of dump trucks, mixers, and dozers were identical.

Experiments were conducted to test the methodology in various cases and to validate the effect of the assistant generator. The structure of the classifier designed for the experiments is shown in Figure 4. Because the goal of the experiments was not to achieve the highest accuracy but to observe the relative accuracy differences in various cases, a simple classifier was created using convolution layers and fully connected layers. To evaluate the performance of the classifier, 240 validation images were randomly selected in advance. In each case, the classifier was trained with 10000 iterations, and the batch size was 32. The best validation accuracy during the training was recorded. For the reliability of the results, the training process was repeated for 50 times in each case. The average of the 50 best validation accuracy values was set as the representative value of the case.

To validate the effect of the assistant generator, preliminary experiments were conducted, as shown in Table 1. For 120 training samples of 32-32-3 (case 1), 64-64-3 (case 2), and 128-128-3 images (case 3), the classifier was evaluated in the following three conditions: 1) using only the original samples, 2) using the augmented data without the assistant generator, and 3) using the augmented data with the assistant generator.

**Table 1** Effect of the assistant generator.

| Case | Performance | Original | Augmented *without* assistant generator | Augmented *with* assistant generator |
|---|---|---|---|---|
| 32-32-3 / 120 Samples | Mean (%) | 61.89 | 68.92 | 69.06 |
| | Std (%) | 2.042 | 1.395 | 1.368 |
| 64-64-3 / 120 Samples | Mean (%) | 64.72 | 70.69 | 71.65 |
| | Std (%) | 1.938 | 1.565 | 1.741 |
| 128-128-3 / 120 Samples | Mean (%) | 66.84 | 71.40 | 71.41 |
| | Std (%) | 1.764 | 1.679 | 1.607 |

**Table 2** Performance of the classifier in nine cases.

| Case | Image size | Training samples | Original performance (%) | Augmented performance (%) | Performance difference (%) |
|---|---|---|---|---|---|
| 1 | 128-128-3 | 480 | 78.42 | 82.71 | 4.290 |
| 2 | 128-128-3 | 240 | 71.72 | 74.63 | 2.916 |
| 3 | 128-128-3 | 120 | 66.84 | 71.41 | 4.570 |
| 4 | 64-64-3 | 480 | 78.45 | 82.04 | 3.590 |
| 5 | 64-64-3 | 240 | 70.72 | 74.23 | 3.510 |
| 6 | 64-64-3 | 120 | 64.72 | 71.65 | 6.934 |
| 7 | 32-32-3 | 480 | 77.56 | 79.21 | 1.645 |
| 8 | 32-32-3 | 240 | 71.33 | 73.55 | 2.221 |
| 9 | 32-32-3 | 120 | 61.89 | 69.06 | 7.170 |
| Average | | | | | 4.094 |

As shown in Table 2, the proposed methodology was evaluated with nine different examples. The number of training samples and the size of the input images were adjusted. For each case, the performance of the classifier was monitored with and without the proposed data augmentation.

**4.2 Evaluation and discussion**

The effect of the assistant generator was validated by comparing the performances of the classifier with and without the assistant generator. As shown in Table 1, the proposed methodology improved the performance of the classifier in the two conditions: with and without the assistant generator. The use of the assistant generator further improved the classification performance. The assistant generator supported the classifier for better performance, although the degree of improvement was smaller in case 3 than in cases 1 and 2. The results of Table 1 indicate that the assistant generator, owing to its intrinsic nature to keep the best weights up to the training moment, tended to improve the performance of the classifier.

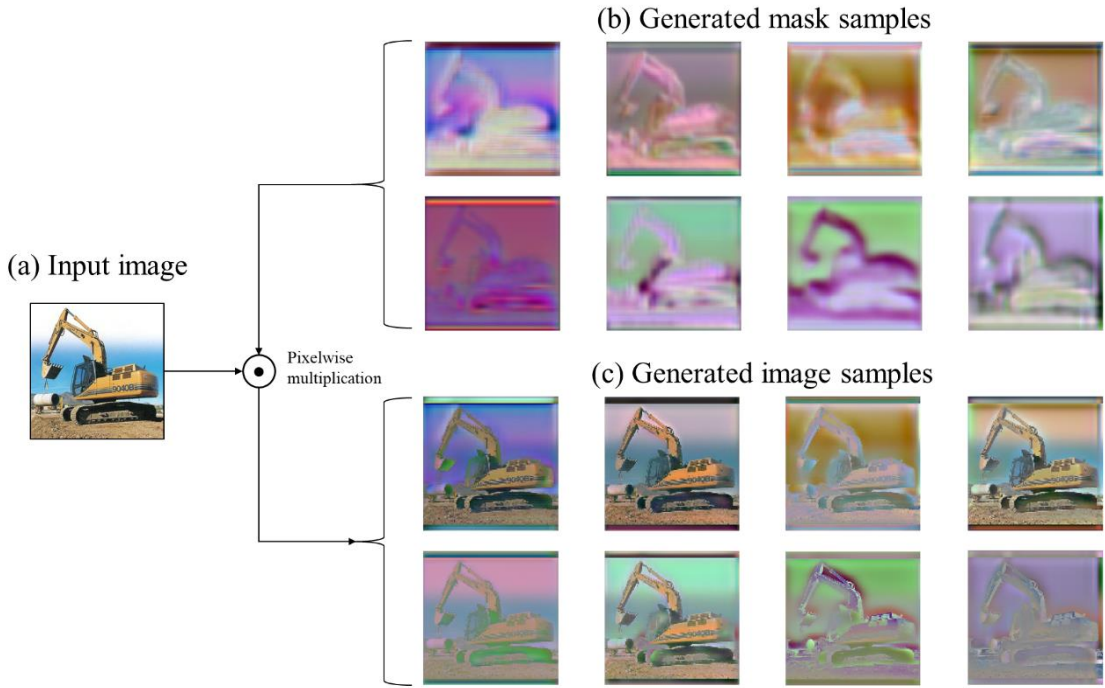

**Figure 5** Sample results of an input image: (a) a sample image from the ImageNet dataset; (b) mask samples generated using input image (a); and (c) image samples generated as results of the pixelwise multiplication of the input image and generated masks.

Table 2 presents the performance of the classifier in the nine different conditions. When the adversarial training was used with the augmented data, the performance of the classifier was improved in all cases. On average, the accuracy was improved by 4.094%. The improvement indicates that the proposed methodology enhanced the classification boundaries in a more robust way with the adversarial samples.

The experimental results indicate that the proposed methodology is more effective for improving the classification performance with relatively small amounts of data. For all the image sizes (128-128-3, 64-64-3, and 32-32-3), the largest improvement in the classification performance was achieved when 120 training samples were used for the data augmentation, compared with the cases where 240 and 480 training samples were used. For cases 3, 6, and 9, the performances of 128-128-3, 64-64-3, and 32-32-3 images, with the use of 120 training samples, improved by 4.570%, 6.934%, and 7.170%, respectively. Additionally, Table 2 indicates that data augmented with smaller numbers of original training samples were sometimes yielded more accurate results than data augmented with larger numbers of original training samples. With 240 training samples of 64-64-3 images, the classification performance based on the original data was 70.72%. However, for the same image size of 64-64-3, the performance based on 120 original training samples and the augmented data was 71.65%, indicating an improved accuracy compared with the case of 240 training samples. These results indicate that the proper use of the methodology can effectively augment data and reduce the effort and time needed for collecting new data.

Sample images generated via the proposed augmentation method are presented in Figure 5. As shown in Figure 5b, the generator of the proposed methodology creates various masks that can apply different values to the foreground, border, and background of an object. Figure 5c shows that the pixelwise multiplication of the masks and the input image preserves the contents of the image while varying the color space. Therefore, the results of this experiment show that a robust augmentation effect is obtained by changing the color space of images. The methodology finds out the effective masks by considering the entire dataset, including both positive and negative training samples. Figure 6 shows that the mask images can vaguely reveal the approximate shapes of the original objects for both positive and negative samples.

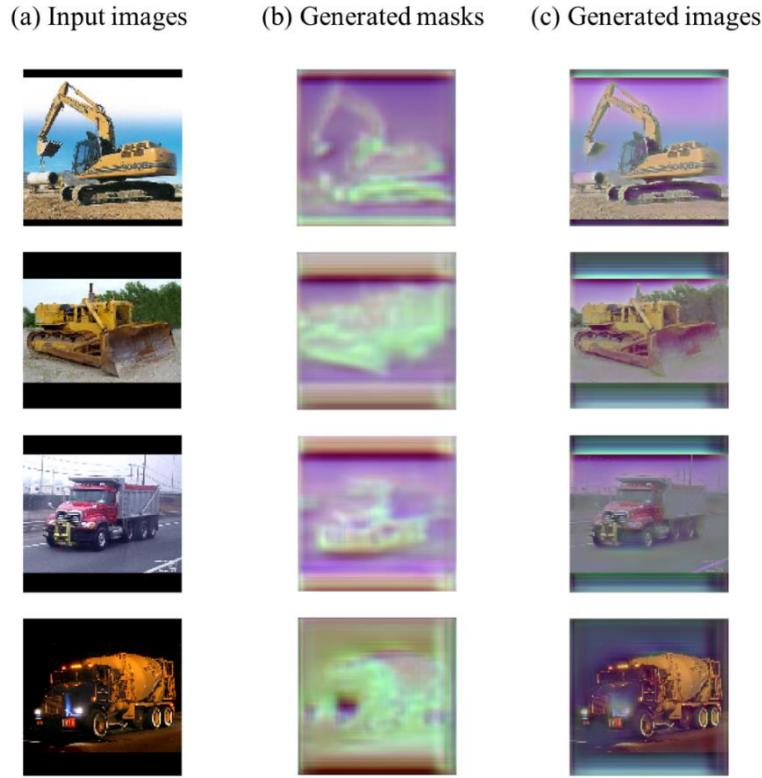

**Figure 6** Sample results of different images: (a) sample input images of an excavator, dozer, dump truck, and mixer from the ImageNet dataset; (b) mask samples generated from the input images; and (c) image samples generated via pixelwise multiplication of the input images and the generated masks.

The effects of the image sizes and number of samples on the classification performance are presented in Table 2. The proposed methodology creates a new image by transforming each pixel of an image. Thus, the number of pixels, resolution, and number of channels of an image indicates the "augmentation space," where various degrees of augmentation are determined. Naturally, a dataset with larger images has a higher capacity to be augmented. This is clearly revealed by a comparison of cases 1, 4, and 7 in Table 2. With image sizes of 32-32-3, 64-64-3, and 128-128-3, the performance improvement after data augmentation with 480 original samples was 1.645%, 3.590%, and 4.290%, respectively. The improvement in the classification performance increased with the image size. This confirms the concept of the "augmentation space," indicating that larger image sizes allow a larger performance improvement.

However, the expectation of the "augmentation space" was not met when the number of training samples was relatively smaller, as revealed by a comparison of cases 3, 6, and 9 in Table 2. With image sizes of 32-32-3, 64-64-3, and 128-128-3, the performance improvement after data augmentation with 120 original samples was 7.170%, 6.934%, and 4.570%, respectively. In contrast to the previous observation, the performance improvement decreased as the image size increased. The stark difference in the improvement patterns is attributed to the number of training samples. With a sufficiently large amount of training samples, a dataset with a larger image size has greater potential to be well augmented. However, when the number of training samples is small, a dataset with a smaller image size has the potential to be well augmented. A large image can be likened to data with a large dimension. A higher dimension corresponds to a larger amount of data needed to train the model. In the cases of 32-32-3 images, the classifier may have already learned enough information from the original training samples, because of the relatively low-dimensional information. If the existing data already provided sufficient information, it may be difficult to find new information through data augmentation. Therefore, as the number of training samples increased, the performance improvement decreased. In contrast, for 64-64-3 and 128-128-3 images, the improvement due to the proposed methodology increased with the number of training samples. It is considered that the parameters of the classifier were still too numerous to be properly trained with the data augmentation.

## 5 CONCLUSION

We propose an adversarial training methodology for data augmentation in construction-equipment classification. The proposed adversarial structure consists of four elements: a generator, assistant generator, discriminator, and classifier. By performing pixelwise multiplication, the generator intentionally creates adversarial samples, which can reduce the classification confidence of the classifier. The performance of the classifier can be improved by augmenting the original training samples with the adversarial samples. The assistant generator can help the classifier not to lose direction during the training, improving the classification performance. The methodology alternates the process of generating adversarial samples and the process of training the classifier in an end-to-end manner.

Experiments confirmed that the assistant generator could improve the classifier performance. The assistant generator further improved the classification performance compared with the cases without the assistant generator. The proposed methodology was validated using nine different cases with 120, 240, and 480 training samples for image sizes of 32-32-3, 64-64-3, and 128-128-3. The experimental results indicated that the methodology achieved an average of accuracy improvement of 4.094%. The performance improvement is expected to be larger when a smaller amount of training samples is used.

The contributions of this study are twofold. First, this is the first attempt to use the GAN-based adversarial training methodology for data augmentation in the construction industry. The proposed method was proven effective for augmenting a dataset of construction-equipment images. Second, the concept of the assistant generator was presented as a part of the proposed method. Selecting the best weights of the assistant generator ensured the optimal classification performance.

The proposed methodology has limitations to be addressed in the future. It does not change the geometric characteristics of the objects, because it changes the color space of a given image through pixelwise multiplication. A future goal would be to generate new information with greater diversity from the given data, even including changes in the geometrical properties. Mixing with other augmentation methods that can give more variations in style can also be performed for generating new information. On the basis of the promising classification-performance improvement, developing a model capable of detection and segmentation may also be a future direction of this study.

## ACKNOWLEDGMENTS

This work was supported by National Research Foundation of Korea (NRF) grants funded by the Ministry of Science and ICT (No. 2018R1A2B2008600) and the Ministry of Education (No. 2018R1A6A1A08025348).